\documentclass{aa_old}
\usepackage{psfig}

\def\ltsima{$\; \buildrel < \over \sim \;$}
\def\lsim{\lower.5ex\hbox{\ltsima}}
\def\gtsima{$\; \buildrel > \over \sim \;$}
\def\gsim{\lower.5ex\hbox{\gtsima}}

\begin{document}

\title{IGR J16194$-$2810: a new symbiotic X--ray binary\thanks{Partly 
based on X--ray observations with {\it INTEGRAL}, an ESA project
with instruments and science data centre funded by ESA member states 
(especially the PI countries: Denmark, France, Germany, Italy, 
Switzerland, Spain), Czech Republic and Poland, and with the participation 
of Russia and the USA, and on optical observations collected at SAAO, 
South Africa.}}

\author{N. Masetti\inst{1},
R. Landi\inst{1},
M.L. Pretorius\inst{2},
V. Sguera\inst{2},
A.J. Bird\inst{2},
M. Perri\inst{3},
P.A. Charles\inst{4},
J.A. Kennea\inst{5},
A. Malizia\inst{1} and
P. Ubertini\inst{6}
}

\institute{
INAF - Istituto di Astrofisica Spaziale e Fisica Cosmica di Bologna, 
via Gobetti 101, I-40129 Bologna, Italy
\and
School of Physics \& Astronomy, University of Southampton, Southampton,
Hampshire, SO17 1BJ, United Kingdom
\and
ASI Science Data Center, via Galileo Galilei, I-00044 Frascati, Italy
\and
South African Astronomical Observatory, P.O. Box 9, Observatory 7935,
South Africa
\and
Pennsylvania State University, 525 Davey Laboratory, University Park, PA 
16802, USA
\and
INAF -- Istituto di Astrofisica Spaziale e Fisica Cosmica di
Roma, Via Fosso del Cavaliere 100, I-00133 Roma, Italy
}

\titlerunning{IGR J16194$-$2810, a new symbiotic X--ray binary}
\authorrunning{Masetti et al.}

\offprints{N. Masetti, {\tt masetti@iasfbo.inaf.it}}

\date{Received 19 March 2007; Accepted 24 April 2007}

\abstract{
We here report on the multiwavelength study which led us to the 
identification of X--ray source IGR J16194$-$2810 as a new Symbiotic 
X--ray Binary (SyXB), that is, a rare type of Low Mass X--ray Binary 
(LMXB) composed of a M-type giant and a compact object. Using the 
accurate X--ray position allowed by {\it Swift}/XRT data, we pinpointed 
the optical counterpart, a M2\,III star. Besides, the combined use of the 
spectral information afforded by XRT and {\it INTEGRAL}/IBIS shows that 
the 0.5--200 keV spectrum of this source can be described with an absorbed 
Comptonization model, usually found in LMXBs and, in particular, in SyXBs. 
No long-term (days to months) periodicities are detected in the IBIS data.
The time coverage afforded by XRT reveals shot-noise variability typical 
of accreting Galactic X--ray sources, but is not good enough to explore
the presence of X--ray short-term (seconds to hours) oscillations in detail.
By using the above information, we infer important parameters for this 
source such as its distance ($\sim$3.7 kpc) and X--ray luminosity 
($\sim$1.4$\times$10$^{35}$ erg s$^{-1}$ in the 0.5--200 keV band), 
and we give a description for this system (typical of SyXBs) in which a 
compact object (possibly a neutron star) accretes from the wind of its 
M-type giant companion. We also draw some comparisons between IGR 
J16194$-$2810 and other sources belonging to this subclass, finding that 
this object resembles SyXBs 4U 1700+24 and 4U 1954+31.
}

\maketitle

\section{Introduction}

Low-mass X--ray Binaries (LMXBs) are interacting systems composed of an 
accreting compact object and a low-mass (1 M$_\odot$ or less) 
main-sequence or slightly evolved late-type star. Recently, a small but 
growing subclass of LMXBs is gaining more attention. The systems belonging 
to this subclass have a M-type giant, rather than a dwarf, as mass 
donor. By analogy with symbiotic stars, in which a white dwarf accretes 
from the wind of a M-type giant companion, they are called symbiotic 
X--ray binaries (SyXBs; Masetti et al. 2006a).

SyXBs are extremely rare: up to now, among more than 150 LMXBs known in 
the Galaxy (Liu et al. 2001), only 4 firm SyXB cases are known: GX 1+4 
(Davidsen et al. 1977; Chakrabarty \& Roche 1997), 4U 1700+24 (Garcia et 
al. 1983; Masetti et al. 2002), 4U 1954+31 (Masetti et al. 2006a,b; 
Mattana et al. 2006) and 
Sct X-1 (Kaplan et al. 2007). All of these objects but GX 1+4 are 
characterized by X--ray emission ranging between $\sim$10$^{32}$ and 
$\sim$10$^{34}$ erg s$^{-1}$, and by the absence of the features typical 
of accreting systems in their optical spectrum (e.g., Masetti et al. 2002, 
2006a); GX 1+4 shows instead a more intense X--ray luminosity (around 
10$^{36}$--10$^{37}$ erg s$^{-1}$) along with a composite optical spectrum 
with strong emission lines (e.g., Chakrabarty \& Roche 1997). All sources 
show long- and short-term X--ray variability.

This difference in the optical spectra is due to the fact that the 
luminosity of a M-type giant is $\sim$10$^{36}$ erg s$^{-1}$ (most of 
which is emitted in the optical and near-infrared ranges), thus only 
in case of large X--ray luminosities can spectral features produced 
by accretion emerge in the optical spectrum. It is thought that this 
difference stems from the mass accretion rate and therefore from the 
evolution of the system, GX 1+4 being likely tighter and more 
evolved according to Gaudenzi \& Polcaro (1999).

All of these systems are suspected to host a neutron star (NS). However, 
only for sources GX 1+4 and Sct X-1, and possibly for 4U 1954+31, is the 
nature of the accretor known: their X--ray emission is pulsed, indicating 
that the accreting object is indeed a NS (Lewin et al. 1971; Koyama et al. 
1991; Corbet et al. 2006, 2007). The system 4U 1700+24 only displays random 
X--ray variability: however, this is more likely due to geometric effects 
rather than to a different type of accretor (Masetti et al. 2002).

Here we present the discovery of the fifth member of the SyXB subclass: 
source IGR J16194$-$2810. This source was first detected in hard 
X--rays above 20 keV with {\it INTEGRAL}, in the 2$^{\rm nd}$ IBIS survey 
(Bird et al. 2006; see also the 3$^{\rm rd}$ IBIS survey of Bird et al.
2007) and in the IBIS extragalactic survey of Bassani et al. (2006), 
at a 20--100 keV flux of $\sim$3$\times$10$^{-11}$ erg cm$^{-2}$ 
s$^{-1}$, assuming a Crab-like spectrum. Through positional cross-correlation 
analysis, Stephen et al. (2006) associated this emission with the {\it ROSAT} 
source 1RXS J161933.6$-$280736 (Voges et al. 1999), which has a flux of 
1.1$\times$10$^{-11}$ erg cm$^{-2}$ s$^{-1}$ in the 0.1--2.4 keV band, 
again assuming a Crab-like spectrum. On statistical grounds, Stephen et 
al. (2006) pointed out that this source is most likely the soft X--ray 
counterpart of IGR J16194$-$2810.

The 8$''$-radius {\it ROSAT} X--ray error box encompasses 3 relatively 
bright optical objects (see Fig. 1). In order to better study this source 
in the soft X--ray band, and to reduce its error circle to pinpoint its 
optical counterpart, we performed observations with with the X--Ray 
Telescope (XRT, 0.3--10 keV; Burrows et al. 2006) on board {\it Swift} 
(Gehrels et al. 2004). 
These observations were part of our program of follow-up pointings of {\it 
INTEGRAL} sources at soft X--rays with {\it Swift}/XRT. 

The capabilities of XRT allow the position of an X--ray source to be 
determined with an uncertainty which can be better than 4$''$, and can 
secure a nominal spectral coverage between 0.3 and 10 keV. We also 
collected hard X--ray archival data of IGR J16194$-$2810 with the IBIS 
instument (Ubertini et al. 2003) on board {\it INTEGRAL} (Winkler et al. 
2003). In parallel, we performed optical spectroscopic observations of the 
field of this source at the South African Astronomical Observatory (SAAO).

The paper is structured as follows: Sect. 2 and 3 will present X--ray
and optical observations of IGR J16194$-$2810, respectively; in Sect. 4
the results of this multiwavelength campaign will be given, and in 
Sect. 5 a discussion on them will be presented. Finally, Sect. 6 will
draw the conclusions and will outline possible future work on this source.
Throughout the paper, uncertainties are given at a 90\% confidence level.

\section{X--ray observations}

We observed the field of IGR J16194$-$2810 with XRT onboard {\it Swift} 
twice; both pointings were performed in Photon Counting mode (see Burrows 
et al. 2006 for details on this observing mode). The log of these 
observations is reported in Table 1.

\begin{table}
\caption[]{Log of the {\it Swift}/XRT observations used in this paper.}
\begin{center}
\begin{tabular}{cccc}
\noalign{\smallskip}
\hline
\noalign{\smallskip}
Observation & Start day & Start time & On-source \\
number & & (UT) & time (ks) \\
\noalign{\smallskip}
\hline
\noalign{\smallskip}
1 & 29 Jan 2007 & 03:36:10 & 5.2 \\
2 & 01 Feb 2007 & 18:21:24 & 2.6 \\
\noalign{\smallskip}
\hline
\noalign{\smallskip}
\end{tabular}
\end{center}
\end{table}

\begin{figure}
\psfig{figure=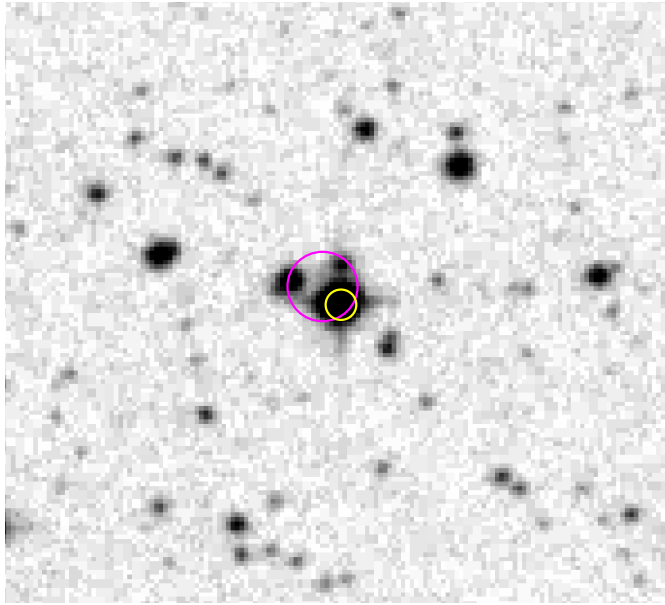,width=8.8cm,angle=0}
\caption[]{DSS-II-Red image of the field of IGR J16194$-$2810 with the 
0.3--10 keV band {\it Swift}/XRT (smaller circle) and the 0.1--2.4 {\it 
ROSAT}/PSPC (larger circle) X--ray positions superimposed. 
The only star positionally consistent with the XRT error circle is
the brighter one at the centre of the image, the M-type giant
USNO-A2.0 U0600\_20227091. In the figure, North is at top and East is to 
the left. The field size is $\sim$2$\farcm$5$\times$2$\farcm$5.}
\end{figure}

The data reduction was performed using the XRTDAS v2.0.1 standard data
pipeline package ({\tt xrtpipeline} v0.10.6) in order to produce the
final cleaned event files.
As in both observations the XRT count rate of the source was high enough 
to produce data pile-up, we extracted the events in an annulus centered on 
the source and 47$''$ wide. The size of the inner circle was determined 
following the procedure described in Romano et al. (2006) and was 7$''$ 
for the first observation and 4$\farcs$7 for the second one.
The source background was measured within a circle with radius
95$''$ located far from the source. The ancillary response file was
generated with the task {\tt xrtmkarf} (v0.5.2) within
FTOOLS\footnote{available at:\\
\texttt{http://heasarc.gsfc.nasa.gov/ftools/}} (Blackburn 1995), and
accounts for both extraction region and PSF pile-up correction. We used
the latest spectral redistribution matrices in the Calibration
Database\footnote{available at:
{\tt http://heasarc.gsfc.nasa.gov/\\docs/heasarc/caldb/caldb\_intro.html}}
(CALDB 2.3) maintained by HEASARC.

We also extracted the spectral and time-series data of this source 
collected with the coded-mask ISGRI detector 
(Lebrun et al. 2003) of the IBIS instrument onboard {\it INTEGRAL}. 
ISGRI data were processed using the standard {\it INTEGRAL}
analysis software (OSA\footnote{available at: \\
{\tt http://isdc.unige.ch/index.cgi?Soft+download}} v5.1; Goldwurm 
et al. 2003); events in the
band 17--300 keV, coming from both fully-coded and partially-coded
observations of the field of view of IGR J16194$-$2810, were 
considered in the analysis. The time resolution for these data was 
that typical of IBIS science windows ($\sim$2 ks). Details on the whole 
procedure can be found in Bird et al. (2007).
Hard X--ray long-term light curves and a time-averaged spectrum were 
then obtained from the available data and using the method described 
in Bird et al. (2006, 2007), for a total of 461 ks on-source collected in 
the time interval October 2002 - April 2006.

\section{Optical observations}

One medium-resolution optical spectrum of the star in the {\it Swift}/XRT 
error box (see Fig. 1 and Sect. 4) was acquired starting at 19:00 UT of 22 
July 2005 with the 1.9-metre ``Radcliffe" telescope located near 
Sutherland, South Africa. The exposure time was 300 s.
This telescope carries a spectrograph mounted at the Cassegrain focus; 
the instrument was equipped with a 1798$\times$266 pixel SITe CCD. 
Grating \#7 and a slit of 1$\farcs$8 were used, providing a 3850--7200 
\AA~nominal spectral coverage. This setup gave a dispersion of 
2.3~\AA/pix.

The spectrum, after correction for flat-field, bias and cosmic-ray 
rejection, was background subtracted and optimally extracted (Horne 
1986) using IRAF\footnote{IRAF is the Image Analysis and Reduction 
Facility made available to the astronomical community by the National 
Optical Astronomy Observatories, which are operated by AURA, Inc., 
under contract with the U.S. National Science Foundation. It is 
available at {\tt http://iraf.noao.edu/}}. 
Wavelength calibration was performed using Cu-Ar lamps, while flux 
calibration was accomplished by using the spectrophotometric standards 
CD $-$32$^\circ$9927 and LTT 377 (Hamuy et al. 1992, 1994). Wavelength 
calibration uncertainty was $\sim$0.5~\AA; this was checked by using the 
positions of background night sky lines.

\begin{figure*}
\vspace{-5cm}
\hspace{-1.5cm}
\psfig{figure=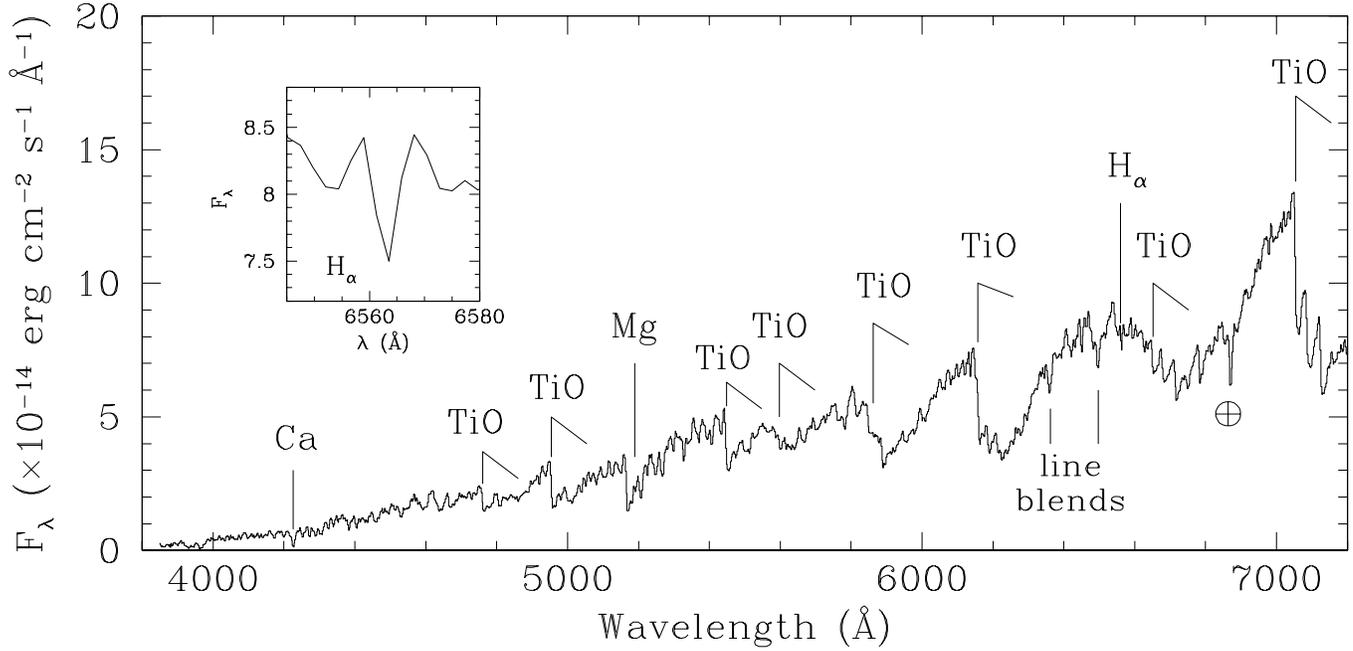,width=21cm,angle=-90}
\vspace{-1.3cm}
\caption[]{3850-7200 \AA~optical spectum of the counterpart of IGR 
J16194$-$2810 obtained with the SAAO 1.9-meter Radcliffe telescope on 
22 July 2005. The spectrum is typical of a star of type M2\,III (see text).
The telluric absorption bands are marked with the symbol $\oplus$.
The inset shows a close-up of the spectrum around the H$_\alpha$ region.}
\end{figure*}

\section{Results}

Only one X--ray source was found in both XRT observations within the 
3$\farcm$5 IBIS error box of IGR J16194$-$2810 (Bird et al. 2007). Using 
the data of XRT obs. 1 (i.e., the deeper one), we determined the position 
of IGR J16194$-$2810 using the {\tt xrtcentroid} (v0.2.7) task. The 
correction for the misalignment between the telescope and the satellite 
optical axis was taken into account (see Moretti et al. 2006 for details). 
The coordinates we obtained for the source are the following (J2000): RA = 
16$^{\rm h}$ 19$^{\rm m}$ 33$\fs$29; Dec = $-$28$^\circ$ 07$'$ 40$\farcs$8 
(with a 90\% confidence level error of 3$\farcs$5 on both coordinates). 
This position is fully consistent with the {\it ROSAT} one (see Fig. 1): 
thus, we can confidently say that these three X--ray objects (the {\it 
INTEGRAL}, the {\it ROSAT} and the {\it Swift} ones) are the same.

Only the brightest of the optical sources within the {\it ROSAT} error 
box, object USNO-A2.0 U0600\_20227091, at coordinates (J2000)
RA = 16$^{\rm h}$ 19$^{\rm m}$ 33$\fs$363; Dec = $-$28$^\circ$ 07$'$ 
39$\farcs$02 (with an error of 0$\farcs$2 on both coordinates:
Deutsch 1999; Assafin et al. 2001), is contained in the XRT uncertainty 
circle, at 2$''$ from the XRT centroid. 

The inspection of the optical 
spectrum of this object (reported in Fig. 2) clearly shows the typical 
features of a M-type star (Jaschek \& Jaschek 1987): it is dominated by 
TiO absorption bands and no emission features typical of X--ray binaries, 
such as Balmer and He {\sc ii} lines, are readily apparent. 
We also find, among the main spectral features, the 
Mg absorption band around 5170 \AA, the Ca {\sc i} line at 4226 \AA~and 
two atomic line blends of metal intersystem lines of Fe {\sc i}, 
Ti {\sc i}, Cr {\sc i}, Ba {\sc i}, Ca {\sc i}, Mn {\sc i}, Co {\sc i} 
and Ni {\sc i} located at 6352 \AA~and 6497 \AA~(see e.g. Turnshek et 
al. 1985). A telluric absorption feature is moreover detected at 
6870 \AA.
A narrow H$_\alpha$ line is detected in absorption, although with 
possible wider emission wings (see inset in Fig. 2), similarly to what 
found by Gaudenzi \& Polcaro (1999) in the optical spectrum of 4U 1700+24. 
However, given that the same profile is seen in the telluric feature 
at 6870 \AA, we believe that this is more due to an effect produced by the 
stellar continuum shape, rather than to the actual presence of emission 
wings around the H$_\alpha$ absorption.

\begin{figure*}
\vspace{-3.7cm}
\hspace{-1.4cm}
\psfig{figure=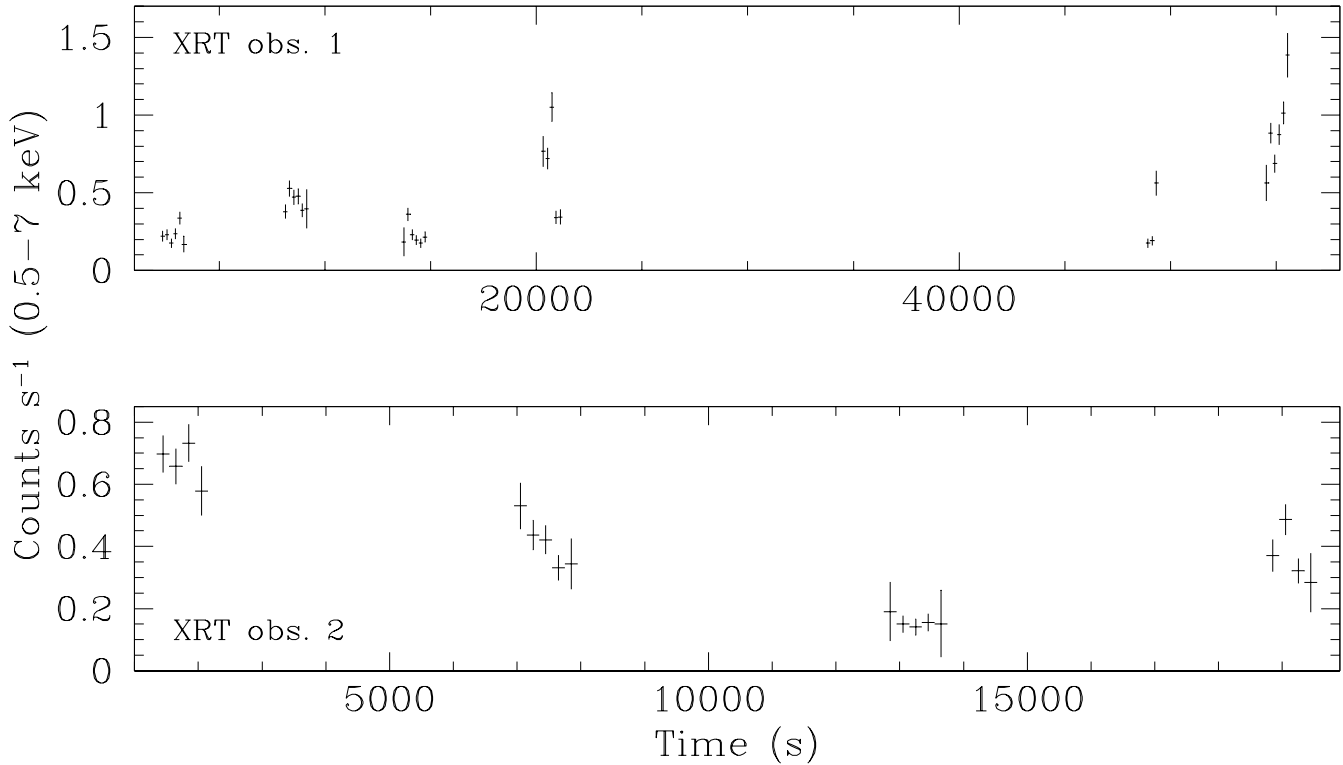,width=21cm,angle=-90}
\vspace{-1.5cm}
\caption[]{0.5--7 keV X--ray light curves, binned at 200 s, of 
IGR J16194$-$2810 as seen during the XRT pointed observations 
reported in the text. Times are in seconds since the beginning 
of the observation as reported in Table 1.}
\end{figure*}

Using the Bruzual-Persson-Gunn-Stryker\footnote{available at:\\ 
{\tt ftp://ftp.stsci.edu/cdbs/cdbs1/grid/bpgs/}} (Gunn \& Stryker 1983) 
and Jacoby-Hunter-Christian\footnote{available at:\\ 
{\tt ftp://ftp.stsci.edu/cdbs/cdbs1/grid/jacobi/}} (Jacoby et al. 1984) 
spectroscopy atlases, we then compared the spectrum of star 
U0600\_20227091 with those of several late-type stars. The best 
match is obtained with star BD $-$02$^\circ$4025 (of type M2\,III), with 
no substantial intervening interstellar absorption. Thus, 
we classify U0600\_20227091 as a star of spectral type M2\,III.

Next, from the $R$-band magnitude ($R\sim$ 11.0) extracted from the 
USNO-A2.0 catalogue\footnote{available at:\\ 
{\tt http://archive.eso.org/skycat/servers/usnoa/}} and from the 
$V-R$ color index of a M2\,III star (1.27; Ducati et al. 2001), we 
determine $V \sim$ 12.3 for the counterpart of IGR J16194$-$2810. 
Assuming that a star of this spectral type has an absolute magnitude 
M$_V$ = $-$0.6 (Lang 1992), we obtain a distance $d \sim$ 3.7 kpc. 
We stress that this should conservatively be considered as an upper limit 
to the distance to this object, as the effect of any amount of 
interstellar absorption along the line of sight was not accounted for.
This correction, however, should not be substantial as the optical 
spectrum of the source shows no evidence of reddening, as mentioned before.

The light curves of the pointed XRT observations (Fig. 3) extracted in the 
0.5--7 keV band, in which the two pointings allowed us to collect 
sufficient statistics from IGR J16194$-$2810, show erratic fluctuations of 
the source emission on variability timescales from hundreds to 
thousands of seconds. This is typical of accreting Galactic sources in 
general and of SyXBs in particular (see Masetti et al. 2002, 2006b). 
The hardness ratio between the 3--7 keV and 0.5--3 keV energy bands did
not change significantly in each of the two single XRT observations, as well
as between them.

Timing analysis on the XRT 0.5--7 keV data was performed with the task 
{\tt powspec} within the FTOOLS package, after having converted the event
arrival times to the Solar System barycentric frame and having considered 
the two {\it Swift} pointing together, so to increase the available 
statistics. We constructed the corresponding Power Spectral Density 
(PSD) using a time resolution of 5 s and dividing the XRT light 
curve in 258 intervals, each one made of 256 bins of 5 s duration.
The PSD thus obtained, in the time frequency interval 
$f$ $\sim$ 10$^{-4}$ -- 0.1 Hz, is characterized by red noise with a 
1/$f$ trend. This is typical of sources showing shot-noise variability 
in their X--ray light curve.

Using the Lomb-Scargle method as described in Sguera et al. (2007), we 
then investigated the 20--100 keV long-term ISGRI light curve of IGR 
J16194$-$2810 to search for periodicities on days to months timescales, 
possibly connected with the orbital period of the system, as in the case 
of 4U 1700+24 (which displays a periodicity of $\approx$400 days: Masetti 
et al. 2002; Galloway et al. 2002). No indication of any periodic 
modulation was found in the range between 1 and $\sim$400 days. A similar 
investigation was performed in narrower spectral ranges (17--30, 20--40 
and 18--60 keV) to search for periodic signals limited to these bands, but 
an identical null result was obtained.

\begin{figure*}
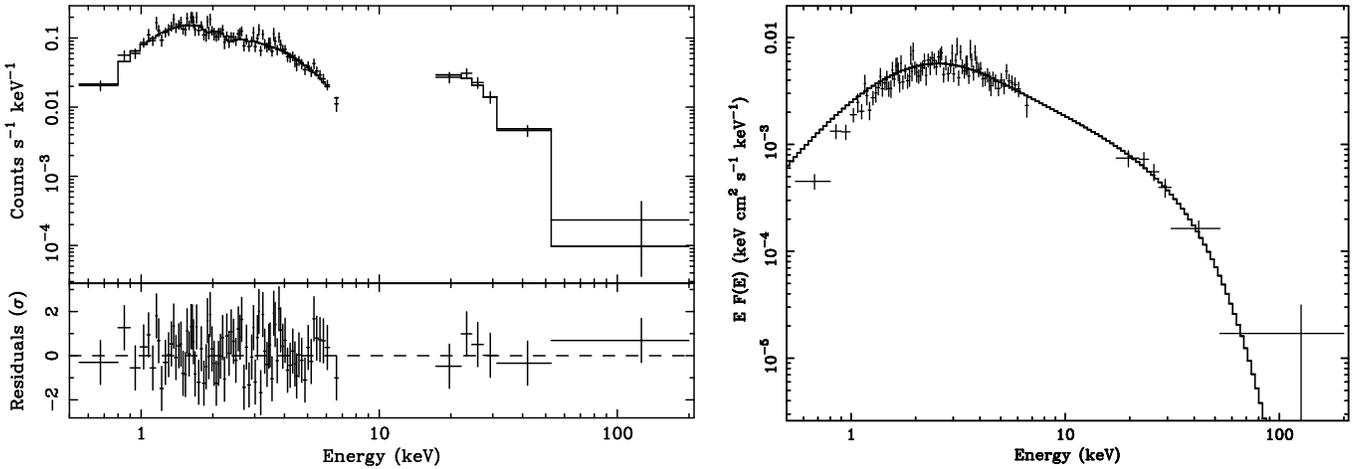

\psfig{figure=7509f4l.ps,width=9.2cm,angle=-90}
\vspace{-6.2cm}
\hspace{9.5cm}
\psfig{figure=7509f4r.ps,width=8.3cm,angle=-90}
\caption[]{{\it Left:} averaged 0.5--200 keV X--ray spectrum of IGR 
J16194$-$2810 obtained from the XRT and ISGRI data described in the text. 
The fit residuals using the best-fit model reported in Table 2 are also 
shown. {\it Right:} deconvolved, absorption-corrected $E$$\times$$F$($E$) 
best-fit model of the X--ray spectrum (continuous line histogram) 
overplotted on the actual X--ray spectral data of the source.}
\end{figure*}

\begin{table}
\caption[]{Best-fit parameters of the Comptonization model adopted 
to describe the 0.5--200 keV X--ray spectrum of IGR J16194-2810
presented in this paper. As quoted in the text, uncertainties 
are given at a 90\% confidence level. The reported fluxes are 
in erg cm$^{-2}$ s$^{-1}$ and are corrected for the intervening 
absorption column as determined from the spectral fitting. The 0.5--200 
keV luminosity, expressed in erg s$^{-1}$, is computed assuming a 
distance $d$ = 3.7 kpc to IGR J16194-2810.}
\vspace{-.2cm}
\begin{center}
\begin{tabular}{lr}
\noalign{\smallskip}
\hline
\noalign{\smallskip}
\multicolumn{1}{c}{Parameter} & \multicolumn{1}{c}{Value} \\
\noalign{\smallskip}
\hline
\noalign{\smallskip}
$\chi^2$/dof & 133/146 \\
$N_{\rm H}$ (10$^{22}$ cm$^{-2}$) & 0.16$^{+0.08}_{-0.07}$ \\
$kT_0$ (keV) & 0.63$^{+0.08}_{-0.07}$ \\
$kT_e$ (keV) & 7.6$^{+6.8}_{-1.6}$ \\
$\tau$ & 6.8$^{+2.3}_{-3.2}$ \\
$K_{\rm Comp}$ & (1.7$^{+0.6}_{-0.9}$)$\times$10$^{-3}$ \\
\noalign{\smallskip}
\hline
\noalign{\smallskip}
$F_{\rm (0.5-2 \,keV)}$   & 7.7$\times$10$^{-12}$ \\
$F_{\rm (2-10 \,keV)}$    & 4.4$\times$10$^{-11}$ \\
$F_{\rm (20-100 \,keV)}$  & 1.6$\times$10$^{-11}$ \\
$F_{\rm (0.5-200 \,keV)}$ & 8.8$\times$10$^{-11}$ \\
\noalign{\smallskip}
\hline
\noalign{\smallskip}
$L_{\rm (0.5-200 \,keV)}$ & 1.4$\times$10$^{35}$ \\
\noalign{\smallskip}
\hline
\noalign{\smallskip}
\end{tabular}
\end{center}
\end{table}

X--ray spectral analysis was performed with the package {\sc xspec} (Arnaud 
1996) v11.3.2. The time-averaged spectra were rebinned to have a minimum of 20 
counts per bin, such that the $\chi^2$ statistics could reliably be used.
As no significant variations in the X--ray hardness of the source 
were found during both {\it Swift} observations, we decided to 
consider the 0.5--7 keV XRT spectrum averaged over the two pointings.
We also accumulated a 18--200 keV ISGRI spectrum averaged over the entire 
on-source time spent by {\it INTEGRAL} on IGR J16194$-$2810. A 
normalization factor between the XRT and the ISGRI spectra was 
introduced to take the non-simultaneity of the observations into account. 

We first attempted a fit of the X--ray spectrum with a simple absorbed 
power law, returning a photon index $\Gamma$ = 1.66$^{+0.12}_{-0.11}$. However, 
the obtained $\chi^2$ is 221 for 152 degrees of freedom (dof). This fact 
and the examination of the fit residuals suggest that a high-energy spectral 
break is present and that a more detailed model is needed to describe the 
X--ray spectrum of the source.

Next, following our past experience on SyXBs (Masetti et al. 2002, 2006b), 
we fit the averaged spectrum of the source with a more physical model 
composed of a thermal Comptonization (Titarchuk 1994) attenuated by a neutral 
hydrogen column. With this model we obtained a satisfactory description of the 
spectrum, as reported in Table 2 and in Fig. 4, in the assumption of a
spherical distribution of the Comptonization plasma around the accreting 
compact object. According to the $F$-test statistics (e.g., Bevington 1969),
the chance improvement probability of this model over the simple power-law 
description is 2$\times$10$^{-3}$, indicating the better statistical quality
of the Comptonization model for the X--ray spectrum of IGR J16194$-$2810.

The use of a disk geometry for the Comptonization plasma gives parameter 
values which are consistent within errors with those obtained assuming a 
spherical distribution but $\tau$, which in this case is 3.2$^{+1.1}_{-1.8}$ 
and therefore only marginally compatible with the value of this parameter in 
the spherical case. However (see also next Section), due to the fact that 
the accretion flow in this system is likely stemming from the stellar 
wind of the secondary (which carries little angular momentum), it is 
fair to assume that the accretion geometry has a spherical form.

It was also found that the XRT/ISGRI intercalibration factor is of order 
unity ($\sim$1.5), indicating that the source did not undergo severe changes 
in the intensity between the {\it INTEGRAL} and the {\it Swift} observations.

No iron emission was found in the spectrum: assuming a Fe line with 
energy 6.7 keV and a width of 0.1 keV, the 90\% upper limit on its 
equivalent width is 82.5 eV. This result is compatible with those from 
other SyXBs (Masetti et al. 2002, 2006b).

\section{Discussion}

\begin{table*}
\caption[]{Synoptic table containing the main parameters of the 5 SyXBs 
known. The X--ray luminosity $L_{\rm X}$ considered in the table refers to 
the 2--10 keV band.
For the computation of the $L_{\rm X}$/$L_{\rm secondary}$ ratios,
bolometric luminosities of the secondary stars are taken from Lang (1992).
The mass accretion rate $\dot{M}$ was computed assuming an accreting NS 
with radius $R_{\rm NS}$ = 10 km and mass $M_{\rm NS}$ = 1.4 $M_\odot$.}
\begin{center}
\begin{tabular}{l|ccccc}
\noalign{\smallskip}
\hline
\hline
\noalign{\smallskip}
\multicolumn{1}{c|}{Parameter} & GX 1+4 & 4U 1700+24 & 4U 1954+31 & Sct X-1 & IGR J16194$-$2810 \\
\noalign{\smallskip}
\hline
\noalign{\smallskip}
Spectral type & & & & & \\
of the secondary & M5\,III [1] & M2\,III [2] & M4-5\,III [3] &  M0\,I ? [4] & M2\,III [5] \\
 & & & & & \\
$V$-band magnitude & 18.4 [1] & 7.7 [6] & 10.7 [3] & 6.6 ($K_s$) [4] & 12.3 [5] \\
 & & & & & \\
$A_V$ (mag) & 5.0 [1] & $\approx$0 [2] & $\approx$0 [3] & $\approx$24 [4] & $\approx$0 [5] \\
 & & & & & \\
Distance (kpc) & 3--6 [1] & 0.42 [2] & $\la$1.7 [3] & $\approx$4 ? [4] & $\la$3.7 [5] \\
 & & & & & \\
X--ray spectrum & [7] & [2,8] & [9]  & [4,10] & [5] \\
 & & & & & \\
$L_{\rm X}$ (erg s$^{-1}$) & $\sim$10$^{37}$ [1] & 
	2$\times$10$^{32}$--10$^{34}$ [2] & 4$\times$10$^{32}$--10$^{35}$ [3,9] & 
	$\approx$2$\times$10$^{34}$ ? [4] & $\la$7$\times$10$^{34}$ [5] \\
 & & & & & \\
$L_{\rm X}$/$L_{\rm secondary}$ & $\sim$2.9 & 10$^{-4}$ -- 5$\times$10$^{-3}$ &
	1.1$\times$10$^{-4}$ -- 3$\times$10$^{-2}$ & 1.3$\times$10$^{-4}$ ? &
	$\la$3.3$\times$10$^{-2}$ \\
 & & & & & \\
$\dot{M}$ (g s$^{-1}$) & 5.4$\times$10$^{16}$ & 
	1.1$\times$10$^{12}$ -- 5.4$\times$10$^{13}$ & 
	2.2$\times$10$^{12}$ -- 1.1$\times$10$^{15}$ & 
	$\approx$1.1$\times$10$^{14}$ ? & $\la$3.8$\times$10$^{14}$ \\
 & & & & & \\
$P_{\rm spin}$ (s) & $\sim$140 [7] & --- & $\sim$18300 [11] & 113 [4] & --- \\
 & & & & & \\
$\dot{P}_{\rm spin}$/$P_{\rm spin}$ (s$^{-1}$) & variable [7, 12] & --- & 
	$-$1.4$\times$10$^{-9}$ [11] & 3.9$\times$10$^{-9}$ [4] & --- \\
 & & & & & \\
$P_{\rm orb}$ (d) & 304 [13, 14] & 404 [2,15] & --- & --- & --- \\
\noalign{\smallskip}
\hline
\noalign{\smallskip}
\multicolumn{6}{l}{References: [1] Chakrabarty \& Roche (1997); [2] Masetti et al. (2002);
[3] Masetti et al. (2006a);} \\
\multicolumn{6}{l}{[4] Kaplan et al. (2007); [5] this work; [6] Tomasella et al. (1997); 
[7] Ferrigno et al. (2007);} \\
\multicolumn{6}{l}{[8] Tiengo et al. (2005); [9] Masetti et al. (2006b); 
[10] Cooke et al. (1984); [11] Corbet et al. (2006);} \\ 
\multicolumn{6}{l}{[12] Nagase (1989); [13] Cutler et al. (1986); [14] Pereira et al. 
(1999); [15] Galloway et al. (2002).} \\
\noalign{\smallskip}
\hline
\hline
\noalign{\smallskip}
\end{tabular}
\end{center}
\end{table*}

The X--ray spectral evidence, as well as the X--ray light curve behaviour 
of IGR J16194$-$2810 strongly suggests that this source is a Galactic 
X--ray binary. Furthermore, the positional coincidence of this X--ray 
source with a M-type giant, as a matter of facts, indicates that the two 
objects are likely the same and that IGR J16194$-$2810 is a SyXB.
For the sake of comparison, and for the reader's use, we collect 
in Table 3 the main properties of the 5 SyXBs known up to now, thus
extending Table 2 of Galloway et al. (2002) with the new results on 
the sources belonging to this subclass of LMXBs.

Following Kaplan et al. (2007), one can use the 2MASS $K$-band 
near-infrared magnitude of star U0600\_20227091 (Skrutskie et al. 2006) to 
evaluate the chance coincidence probability of finding a bright red giant 
within the XRT error box. This star has a magnitude $K$ = 6.98; in this 
area of the sky, and the number of stars brighter than $K$ = 7 mag 
is $\sim$15 per square degree. This means that the chance probability of 
finding such a bright star within the XRT error circle is 
$\sim$4$\times$10$^{-5}$.
Therefore, the low random chance of a positional coincidence of this
X--ray source with a M-type giant further strongly supports the physical 
association of these two objects.

Thus, we can confidently state that U0600\_20227091 is indeed the actual 
optical counterpart of IGR J16194$-$2810, and that this X--ray object is 
indeed a SyXB located at $\la$3.7 kpc from Earth. This distance is similar 
to that suggested by Kaplan et al. (2007) for the SyXB Sct X-1. It also 
means that the 2--10 keV band luminosity of this source is 
$\la$7.2$\times$10$^{34}$ erg s$^{-1}$, which is similar to that of other 
objects of this subclass (e.g., Masetti et al. 2002, 2006a,b; see also 
Table 3). If we compare the 2--10 keV X--ray luminosity of the system with 
the total luminosity of its M-type giant companion, $\sim$550 $L_\odot$ 
(Lang 1992), i.e. 2$\times$10$^{36}$ erg s$^{-1}$ (most of which is 
emitted in the optical and near-infrared bands) we see that, as for nearly 
all other SyXBs, the optical light due to the reprocessing of X--ray 
irradiation is overwhelmed by the emission of the M-type giant star.

We note that the absence of apparent interstellar absorption in the 
optical spectrum is at odds with the $N_{\rm H}$ obtained from our X--ray 
spectral fitting (see Table 2), which implies $A_V \sim$ 1 mag, according to 
the empirical formula of Predehl \& Schmitt (1995). This suggests that most 
of this hydrogen column, likely connected with the accretion stream, is 
concentrated around the compact object. This is not uncommon in SyXBs 
(see e.g. Masetti et al. 2006b).

The PSD obtained from the XRT data has the characteristics of the 1/$f$-type
shot-noise variability often seen in this class of objects (e.g., Masetti et 
al. 2002, 2006b) and likely due to random instabilities in the accretion 
process, or to inhomogeneities in the accreting stellar wind captured 
by a NS (see e.g. Kaper et al. 1993). Thus, a straightforward explanation 
for the X--ray activity from this source is that it is produced by
inhomogeneities in the accretion flow onto a compact object, possibly a NS 
(e.g. van der Klis 1995).

Although no short-term periodicity linked to the spin of the accreting 
object has been found in the XRT data analysis, it may be likely that, if the 
accretor is a NS, this spin period is $\la$100 s.
Indeed (see also Table 3), we note that slowly rotating pulsars appear 
to be usual in SyXBs, as the NSs hosted in these systems display spin 
periodicities ranging from hundreds of seconds (Lewin et al. 1971; Kaplan 
et al. 2007) to hours (Corbet et al. 2006, 2007). 
Thus, any theory aiming at an accurate description of the evolution of 
this kind of LMXB should also explain this peculiarity.

The best-fit X--ray spectral model is also typical of X--ray binary 
systems hosting a NS accreting from a stellar wind (e.g., Masetti 
et al. 2004, 2006c).
We should note that the presence of a black hole (BH), rather than a NS, 
in this system cannot however be excluded; nevertheless, the temperatures 
associated with the Comptonization component are those generally seen 
in LMXBs hosting an accreting NS (e.g., Paizis et al. 2006).

A possible further indication that the accreting matter is flowing onto the 
polar caps of a NS via magnetic field confinement comes from the estimate 
of the size $r_0$ of the region emitting the Comptonization soft X--ray seed
photons. Following the prescription by in 't Zand et al. (1999) for the
computation of $r_0$, and using the best-fit spectral parameters reported
in Table 2 in the assumption of a spherical plasma cloud, we obtain that 
$r_0 \sim$ 1.4 km. This estimate suggests that the area emitting soft seed 
photons on the NS covers only a fraction of its surface and it is comparable 
with the size of the base of an accretion column, which is $\approx$0.1 times 
the NS radius (e.g., Hickox et al. 2004).

The fact that, assuming that an accreting NS is harboured in this system, 
IGR J16194$-$2810 does not show X--ray pulsations, at variance with 
GX 1+4 or Sct X-1, may be due to geometric effects (such as a low inclination 
of the system and/or the quasi-alignment between the rotation and the magnetic 
field axes), as invoked by Masetti et al. (2002, 2006b) for 4U 1700+24 and 
4U 1954+31. Alternatively, the plasma cloud surrounding the NS may 
completely comptonize the soft X--ray emission coming from it and 
therefore may smear out any periodic modulation emitted by the NS surface 
(e.g., Titarchuk et al. 2002); indeed, this latter interpretation is 
supported by the fact that we do not see any direct thermal radiation from 
the NS in the X--ray spectrum. Of course, we cannot exclude that a 
combination of the two effects above is at work in IGR J16194$-$2810.

We thus suggest that IGR J16194$-$2810 is a SyXB with overall 
characteristics which are broadly similar to those of systems 4U 1700+24 
and 4U 1954+31, in which a compact object, likely a NS, moves around a 
M-type giant in a wide orbit and accretes from its stellar wind (Masetti 
et al. 2002, 2006b).

\section{Conclusions}

Using multiwavelength information extending from optical wavelengths to 
the hard X--ray range we characterized the nature of source IGR 
J16194$-$2810 and found that it is a SyXB possibly hosting a (slowly 
rotating?) NS accreting from the wind of its M-type giant companion. 
This is the fifth object belonging to this small but growing class of 
peculiar Galactic X--ray binaries.

Future optical spectrophotometry studies on this object may shed light on 
the determination of its orbital period; likewise, the study of its 
long-term X--ray behaviour using monitoring instruments at energies below 
20 keV can help in the determination of any long-term periodicity.
In parallel, pointed observation with X--ray satellites with high spectral
(e.g., {\it XMM-Newton} or {\it Chandra}) and temporal ({\it RXTE})
sensitivity can explore the nature of the detected X--ray emission
and can search for the presence of any pulsed signal or of other 
short timescale periodicities in the X--ray light curve of this object, 
for a full characterization of this rare source.

\begin{acknowledgements}

We thank Giancarlo Cusumano for useful advices concerning the XRT data 
reduction and analysis, and Mauro Orlandini, Loredana Bassani and Eliana 
Palazzi for several important remarks and suggestions. We also thank the 
anonymous referee for useful comments which helped us to improve the 
quality of this paper. This research has made use of the NASA's 
Astrophysics Data System, of the 2MASS survey, of the HEASARC archive, and 
of the SIMBAD database operated at CDS, Strasbourg, France.
The authors acknowledge the ASI and INAF financial support via grant 
No. 1/023/05/0.

\end{acknowledgements}

\end{document}